\documentclass{aa}  
\usepackage{CJKutf8}
\usepackage{graphicx}
\usepackage[breaklinks,colorlinks,citecolor=blue,linkcolor=blue]{hyperref}
\usepackage{txfonts}
\usepackage{lscape}
\usepackage{epstopdf}
\usepackage{CJK}
\usepackage{float}
\usepackage[absolute]{textpos}
\usepackage{amsmath}
\usepackage{mathrsfs}

\usepackage{arydshln}

\begin{document} 
   \title{Accelerating star formation of dense clumps}
   \author{Xunchuan Liu \begin{CJK}{UTF8}{gbsn}(刘训川)\end{CJK}
          \inst{1}\thanks{liuxunchuan001@gmail.com}
          } 

   \institute{
   Leiden Observatory, Leiden University, P.O. Box 9513, 2300RA
Leiden, The Netherlands 
              }
              
\date{Received xxx xx}
\abstract{
We present a statistical framework that establishes an accelerating star formation scenario for dense clumps using ATLASGAL and ALMAGAL samples. By employing the cumulative distribution function of dust temperature as a monotonic evolutionary indicator, we linearize clump evolution into a normalized timescale, enabling direct comparison across different samples. The virial mass of clumps increases exponentially with this normalized time, revealing an accelerating buildup of star-forming gas within protoclusters. The evolution of the maximum core mass further shows that the growth timescales of protoclusters and their embedded most massive protostars are comparable, implying a self-similar acceleration of star formation from the stellar to the protocluster scale. This unified framework naturally reproduces the observed evolution of luminosity, the core mass function, the mass growth of the most massive protostars, and the dense gas star formation law on clump scales, establishing a coherent picture of accelerating star formation across scales.
}


 \maketitle
 

\section{Introduction} \label{sec_intro}
Most young stars, especially massive ones, form in protoclusters within dense clumps \citep[e.g.,][]{2016A&A...588A..29H}, following certain universal physical laws, such as the initial mass function \citep[IMF;][]{1955ApJ...121..161S}, the core mass function \citep[CMF;][]{1998A&A...336..150M}, and the star formation law \citep{1998ApJ...498..541K,2004ApJ...606..271G}. Dense clumps (with radii approximately between 0.1 and 1 pc) are therefore believed to serve as the fundamental building blocks of star formation, giving rise to the key physical processes that regulate it \citep{2003ApJ...585..850M,2004MNRAS.349..735B,2025RAA....25b5020L}. Large surveys of dense clumps  \citep[e.g., ATLASGAL;][]{2009A&A...504..415S} and their embedded proto-clusters (e.g., ALMAGAL; \citealp{2025A&A...696A.149M}; ALMA-QUARKS; \citealp{2024RAA....24b5009L})
now provide an opportunity to statistically investigate the physical processes that underpin these fundamental laws of star formation. However, the lack of tracers that can provide a linear evolutionary timescale for individual clumps makes it difficult to constrain how these processes evolve over time.

Fortunately, for a survey containing a large, unbiased sample of sources spanning an evolutionary sequence, if one or more measured parameters vary monotonically with evolution, it is possible to use the cumulative distribution function (CDF) of these parameters as a proxy for a linear evolutionary timescale. This technique has been applied to the infrared spectral energy distribution (SED) slope ($\alpha$) to trace the evolutionary age of protoplanetary disks \citep{2024RAA....24g5001L}. 
Here, we assume that other physical parameters mainly introduce random scatter into the chosen tracer. As long as this scatter is not systematic, it only broadens the distribution and does not prevent the underlying evolutionary trend from being recovered.
In the context of young star and protocluster formation, the ATLASGAL survey \citep{2009A&A...504..415S} provides such an unbiased sample, where the clump dust temperature ($T_{\rm dust}$), which is believed to increase with evolution, may serve as a suitable monotonic parameter. The obtained $T_{\rm dust}$-age relation can also be used to calibrate the evolutionary age of other samples that may be affected by selection biases, such as the ALMAGAL survey \citep{2025A&A...696A.149M}. In this work, we use the CDF of $T_{\rm dust}$ to linearize the evolutionary trend of dense clumps and to uncover the underlying physical laws of star formation that have remained obscured due to the lack of reliable age indicators for interpreting the data.
An exponentially accelerating star formation scenario for dense clumps is proposed.

\section{Data sources}

\begin{figure*}
\centering
\includegraphics[width=0.99\linewidth]{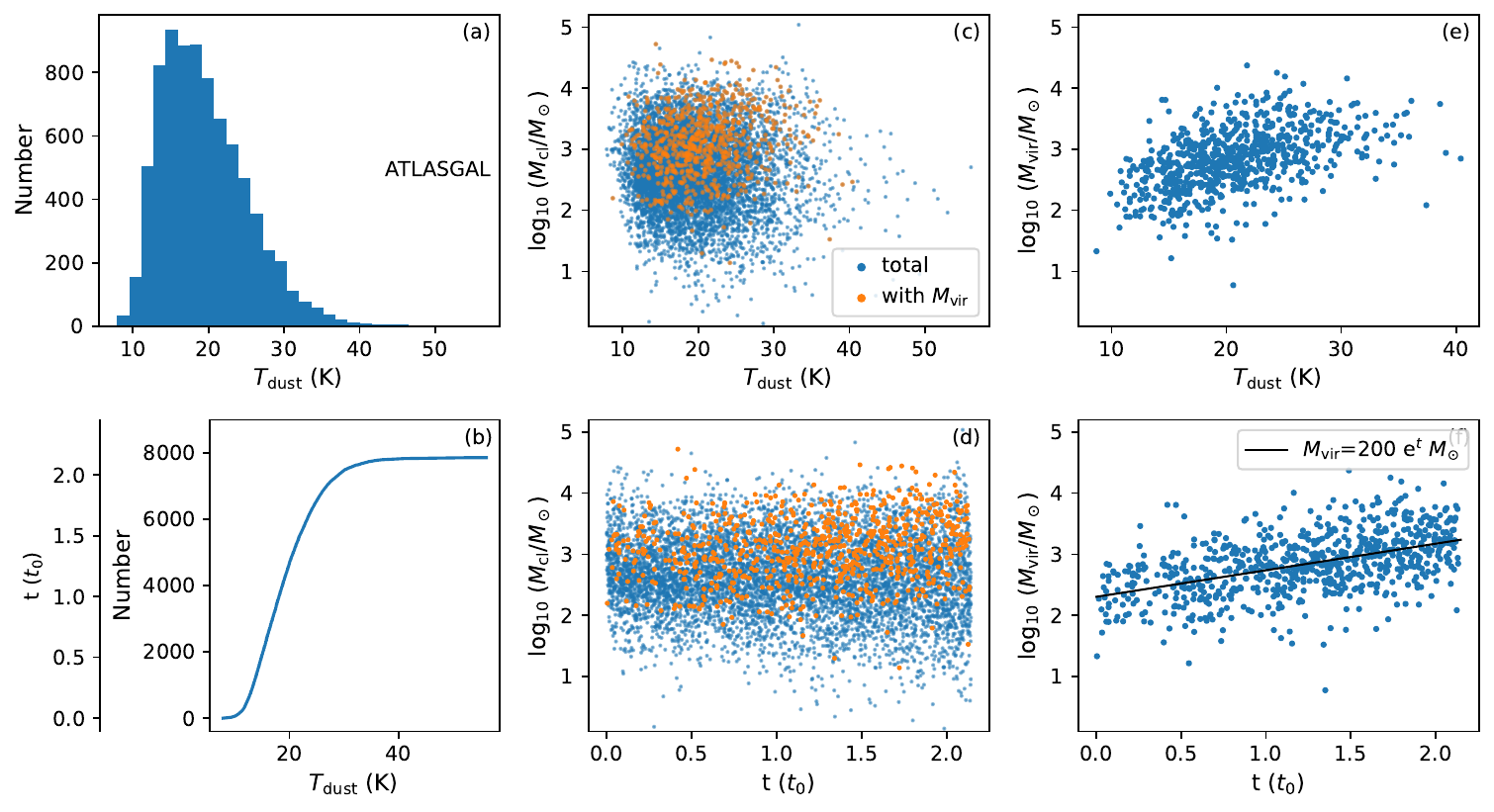}
\caption{
Distribution of ATLASGAL clumps \citep{2018MNRAS.473.1059U}. 
(a) Probability density function (PDF) of dust temperature ($T_{\rm dust}$). 
(b) Cumulative distribution function (CDF) of $T_{\rm dust}$. The two y-axes indicate the linear mapping between the CDF of $T_{\rm dust}$ and the evolutionary age ($t$). 
(c) Distribution of clump mass ($M_{\rm cl}$) versus $T_{\rm dust}$ (blue dots); orange dots represent clumps with virial mass measurements. 
(d) Same as panel (c), but with the x-axis replaced by $t$. 
(e) Distribution of virial mass ($M_{\rm vir}$) versus $T_{\rm dust}$. 
(f) Same as panel (e), but with the x-axis replaced by $t$. 
The evolutionary age $t$ is normalized (in units of $t_0$) such that the exponential fit between $M_{\rm vir}$ and $t$ (bla) follows the relation $M_{\rm vir} \propto e^{t}$.
\label{fig_atlasgal}}
\end{figure*}

The APEX Telescope Large Area Survey of the Galaxy (ATLASGAL; \citealp{2009A&A...504..415S,2018MNRAS.473.1059U}) is a large-scale, unbiased $870\;\mu$m continuum survey of the inner Galactic plane, covering approximately 420~deg$^2$ within the longitude range $-80^\circ < \ell < 60^\circ$ and latitude $|b| < 1.5^\circ$. The survey provides a resolution of $19.2''$ and a typical sensitivity of $\sim 50$~mJy~beam$^{-1}$, enabling the systematic detection of dense molecular clumps that are likely sites of high-mass star and protocluster formation. In \citet{2018MNRAS.473.1059U}, a complete sample of $\sim8\,000$ clumps was compiled, with reliable radial velocities and distances determined for $\sim97\%$ of the sources, allowing the derivation of key physical parameters (mass, luminosity, dust temperature) and classification into evolutionary subsamples.
The dust temperatures ($T_{\rm dust}$) of these clumps range from 7.9 to 56.1~K (see also Figure~\ref{fig_atlasgal}). Emission from the NH$_3$~(1,1) line was measured for $\sim$750 clumps (about 10\% of the total sample), and the virial mass of each clump was derived from the NH$_3$ line widths, resulting in virial parameters generally smaller than two \citep{2018MNRAS.473.1059U}. These data reveal statistically significant trends: dust temperature, luminosity ($L$), and line width all increase with evolutionary stage, while clump mass remains largely independent of evolution, implying that the bulk of the clump mass is in place before star formation begins.
Notably, the $T_{\rm dust}$ distribution shows a smooth trend with no significant discontinuities, indicating that star formation is a continuous process and that the observational stages do not correspond to fundamentally distinct physical mechanisms \citep{2022MNRAS.510.3389U}. This supports treating $T_{\rm dust}$ as a continuous tracer of evolutionary age after proper linearization.

The ALMA Evolutionary study of high-mass star formation (ALMAGAL; \citealp{2025A&A...696A.149M}) is a large ALMA Cycle~7 program that maps over 1000 dense clumps from the ATLASGAL survey with a resolution of $\sim$1000~au in Band~6.
The sample of ALMAGAL covers distances up to 7.5~kpc, dust temperatures from $\sim$8 to $\sim$40~K, and spans evolutionary stages from infrared dark clouds to \ion{H}{ii} regions. 
These clumps are massive ($M_{\rm cl} \gtrsim 500$~M$_\odot$) and exhibit a wide range of luminosity-to-mass ratios ($L/M$) spanning $\sim$4 orders of magnitude. The survey resolves the internal substructures of clumps, enabling systematic studies of fragmentation and core evolution. Note that a core typically has a characteristic size of $\sim0.01$ pc, comparable to the ALMAGAL resolution, and harbors one or more young stellar objects. The further fragmentation and multiplicity of individual cores are beyond the scope of this work.
In the first compact-source catalog \citep{2025A&A...696A.151C}, the core mass function (CMF) is found to evolve with time, flattening and extending toward higher masses as clumps progress through their evolutionary stages (as traced by $L/M$ or $T_{\rm dust}$). This result indicates that the CMF builds up gradually, consistent with a clump-fed scenario of high-mass star formation. However, this trend and its underlying physics are still not fully quantified owing to the absence of a linear tracer of evolutionary age.

\section{Results}
\subsection{Linearization of the evolution trend}
To investigate the time-dependent evolution of dense clumps, we adopt the dust temperature ($T_{\rm dust}$) as a monotonic tracer of evolutionary stage. Figure~\ref{fig_atlasgal} illustrates the statistical distributions of ATLASGAL clumps \citep{2018MNRAS.473.1059U} and the procedure used to assign a normalized evolutionary age. The probability density function (PDF) of $T_{\rm dust}$ (panel a) exhibits a continuous distribution across the sample, supporting its use as an evolutionary indicator. The cumulative distribution function (CDF) of $T_{\rm dust}$ (panel b) provides a mapping to a normalized evolutionary age $t$, expressed in units of a characteristic timescale $t_0$:
\begin{equation}
    t = \gamma\ {\rm CDF}(T_{\rm dust})\, t_0, \label{eq_t}
\end{equation}
where $\gamma$ is a scaling factor chosen so that $\gamma\,{\rm CDF}$ spans a convenient range from 0 to a few. This definition allows each clump to be assigned a relative evolutionary age, while the absolute timescale $t_0$ (approximately $10^5$ years) remains undetermined and can be calibrated in future studies.

\begin{figure}
    \centering
    \includegraphics[width=0.99\linewidth]{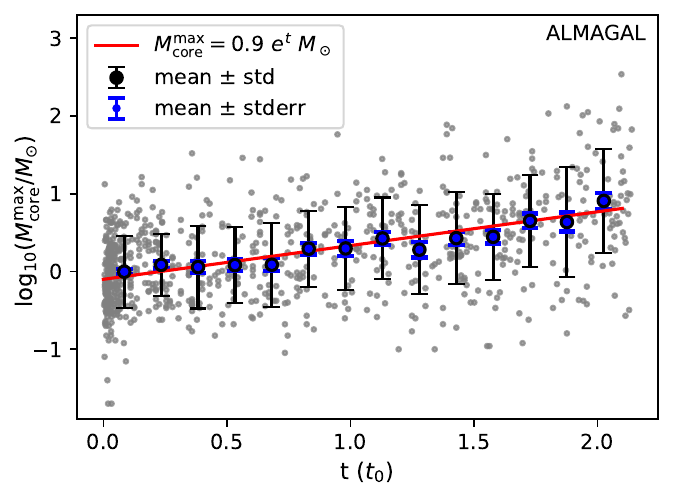}
    \includegraphics[width=0.99\linewidth]{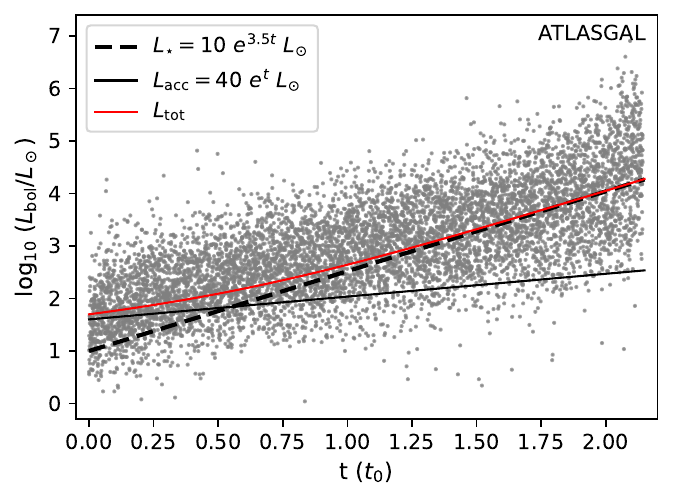}
    \caption{Upper: Distribution of $\log_{10}(M_{\rm core}^{\rm max})$ as a function of $t$ for the ALMAGAL sample.  
Black error bars indicate the mean and standard deviation of $\log_{10}(M_{\rm core}^{\rm max})$ within each time bin, while blue error bars represent the standard error of the mean (i.e., ${\rm stderr}={\rm std}/\sqrt{N}$, with $N$ being the number of sources in each bin).  
The red line shows the exponential fit between $M_{\rm core}^{\rm max}$ and $t$.  
Lower: Distribution of $L_{\rm bol}$ as a function of $t$ for the ATLASGAL sample.  
The red line represents the best-fit model of Eq.~\ref{eq_lumiform}.  
The dashed and solid black lines indicate the contributions from intrinsic stellar luminosity and accretion luminosity, respectively (see Sect.~\ref{sec_lumi}).
    \label{fig_maxcore_L}}
\end{figure}

Using this relative age, we examine how clump properties vary with evolution. Panels (c) and (d) show the distribution of clump mass ($M_{\rm cl}$) as a function of $T_{\rm dust}$ and $t$, respectively, with clumps having virial mass measurements highlighted. The relatively uniform distribution of $M_{\rm cl}$ over $t$ (nearly constant standard deviation) indicates that the ATLASGAL sample is unbiased and consistent with a clump-fed scenario, in which the bulk of the star-forming mass is already present in clumps prior to significant star formation. The subsample with $M_{\rm vir}$ measurements is evenly sampled for $M_{\rm cl} \gtrsim 100$~M$_\odot$, covering the clump mass range relevant for proto-clusters revealed by ALMAGAL
\citep{2025A&A...696A.151C}. The mapping from $T_{\rm dust}$ to $t$ is derived from the full ATLASGAL sample (panel b), ensuring that the calibration of $t$ is representative of the overall population.

Panel (e) displays the distribution of virial mass ($M_{\rm vir}$), which shows a positive correlation with $T_{\rm dust}$. After linearization (panel f), $\log(M_{\rm vir})$ exhibits a tighter linear relation with $t$, indicating an exponential increase of $M_{\rm vir}$ over time. We use this relation to calibrate $\gamma$ such that $M_{\rm vir} \propto e^t$, which for the best fit (black line in panel f) corresponds to
\begin{equation}
M_{\rm vir} = 200\,e^t\,M_\odot.
\end{equation}
Within this framework, $M_{\rm vir}$ can be interpreted as the mass of star-forming material ($M_{\rm SF}$) within the clump, suggesting that clumps undergo exponential accretion of star-forming material as they evolve. The star-forming gas and forming stars together constitute the proto-cluster embedded within the host clump, and $t_0$ can be interpreted as the characteristic timescale for proto-cluster formation, during which the proto-cluster mass increases by a factor of $e$.

\subsection{Maximum core mass}
Cores in dense clumps typically form in protoclusters. Therefore, the total mass of cores within a clump, $M_{\rm protocluster}$, is expected to scale with $M_{\rm SF}$.
The CMF is believed to be closely linked to the IMF \citep{1998A&A...336..150M,2025RAA....25b5020L}, exhibiting a high-mass tail
\begin{equation}
    \frac{dN}{d\ln(M_{\rm core})} \propto M_{\rm core}^{-\alpha},
\end{equation}
with a slope shallower than the Salpeter value ($\alpha_{\rm Salpeter} = 1.35$).
Here, $M_{\rm core}$ denotes the mass of an individual core.
Such top-heavy CMFs are widely observed in massive star formation regions \citep{2018ApJ...853..160C,2021ApJ...921...48S,2024RAA....24b5009L}. We adopt a fiducial high-mass-end CMF with $\alpha = 1$, as suggested by the results of ALMA-IMF \citep{2022A&A...664A..26P,2024A&A...690A..33L}.
The maximum core mass $M_{\rm core}^{\rm max}$ in a protocluster of total mass $M_{\rm protocluster}$ is set by the condition that the cumulative number of cores above $M_{\rm core}^{\rm max}$ is of order unity:
\begin{equation}
    N(>M_{\rm core}^{\rm max}) \sim 1,
\end{equation}
which leads to 
\begin{equation}
    M_{\rm core}^{\rm max} \propto M_{\rm protocluster}^{1/\alpha} \propto M_{\rm protocluster}
    \propto M_{\rm SF}. \label{eq_mcore_mcl}
\end{equation}

The positive correlation between the protocluster mass (represented by the total core mass $M_{\rm core}^{\rm tot}$) and the maximum core mass $M_{\rm core}^{\rm max}$ has been observed in the ALMAGAL sample \citep{2025A&A...696A.151C}. For each host clump in the sample, we calculate the evolutionary age $t$ from the dust temperature $T_{\rm dust}$ (see Eq.~\ref{eq_t} and panel~b of Figure~\ref{fig_atlasgal}). A clear linear relation between $\log_{10}(M_{\rm core}^{\rm max})$ and $t$ is found (upper panel of Figure \ref{fig_maxcore_L}). The best-fit relation is
\begin{equation}
    M_{\rm core}^{\rm max} = 0.9\, e^{t}\, M_\odot. \label{eq_maxcore}
\end{equation}
This implies that
\begin{equation}
    M_{\rm core}^{\rm max} \propto e^{t} \propto M_{\rm SF},
\end{equation}
reproducing the maximum core–cluster mass relation predicted from the CMF (Eq. \ref{eq_mcore_mcl}).

We therefore suggest that both the protocluster mass and the maximum core mass undergo exponentially accelerating mass assembly, characterized by comparable timescales, until $M_{\rm SF}$ approaches the saturation limit set by the available reservoir mass ($M_{\rm cl}$). Furthermore, we assume that all cores follow a similar accretion pattern, such that
\begin{equation}
    \dot{M}_{\rm core} = \frac{d(e^t)}{dt} = e^t = M_{\rm core}. \label{eq_coregrow}
\end{equation}
This accretion pattern is consistent with the concept of competitive accretion, in which more massive cores exhibit higher accretion rates facilitated by the global gravitational potential \citep{2001MNRAS.323..785B}. 

Note that the total core mass represents only a small fraction of $M_{\rm SF}$. Therefore, at any time before saturation, there remains ample mass available both for the formation of new core seeds and for the continued accretion onto existing cores. For simplicity, we neglect the effects of core merging and fragmentation. Assuming a constant initial core mass function (ICMF; the mass distribution of nascent cores), the number of cores ($N_{\rm core}$) and its growth rate ($\dot{N}_{\rm core}$) are both expected to follow an exponential trend:
\begin{equation}
    N_{\rm core} \propto \dot{N}_{\rm core} \propto e^t. \label{eq_Ngrow}
\end{equation}
We will show in Sect.~\ref{sec_cmf} that such a formation and accretion pattern naturally produces the top-heavy fiducial CMF at the high-mass end, independent of the assumed ICMF.

\subsection{Luminosity evolution}\label{sec_lumi}

Massive protostars may ignite nuclear fusion during their main accretion phase; hence both the accretion luminosity ($L_{\rm acc}$) and the intrinsic stellar luminosity ($L_\star$) contribute to the total clump luminosity. 
We adopt a simple model that neglects the modification of $L_\star$ by the accretion rate $\dot M_\star$, and assume the canonical ZAMS mass–luminosity relation
\begin{equation}
    L_\star^{\rm individual} \propto M_\star^{3.5},
\end{equation}
noting that protostellar evolution (e.g., radius inflation at high $\dot M_\star$) can modify this relation \citep[e.g.][]{1991ApJ...375..288P,2010ApJ...721..478H}. 
Under the assumption that the mass of the embedded massive star scales with the maximum core mass, the clump luminosity is dominated by the most massive object,
\begin{equation}
    L_\star^{\rm clump} \propto (M_{\rm core}^{\rm max})^{3.5} \propto e^{3.5t},
\end{equation}
given $M_{\rm core}^{\rm max} \propto e^{t}$ from Eq.~\ref{eq_maxcore}.

The luminosity contributed by accretion  of an individual protostar scales as
\begin{equation}
    L_{\rm acc}^{\rm individual} \sim \frac{G M_\star \dot M_\star}{R_\star},
\end{equation}
where both $R_\star$ and $\dot M_\star$ depend on stellar mass and accretion history. Since $R_\star$ typically increases with $M_\star$, $L_{\rm acc}^{\rm individual}$ depends only weakly on stellar mass. It is therefore reasonable to approximate the protocluster-integrated accretion luminosity as proportional to the global star-forming mass assembly rate:
\begin{equation}
    L_{\rm acc}^{\rm clump} \propto \sum \dot M_\star \propto \dot M_{\rm SF} \propto M_{\rm SF} \propto e^{t}. \label{eq_lumiform}
\end{equation}
Here, we assume that the total mass accreted by all cores is proportional to the growth rate of the star-forming mass within a clump, neglecting any possible time delay between mass growth on different spatial scales.
Thus, while $L_{\rm acc}^{\rm clump}$ grows approximately as $e^{t}$, the intrinsic luminosity of the most massive star increases much more rapidly ($\propto e^{3.5t}$) and soon dominates the thermal feedback once high-mass stars begin to form. 
These relations constitute a simplified, phenomenological framework that assumes a constant core-to-star efficiency and ignores feedback regulation.

The bolometric luminosity of the host clump can then be written as
\begin{equation}
    L_{\rm bol}^{\rm clump} = L_{\rm acc}^{\rm clump} + L_{\star}^{\rm clump}
    = a\,e^{t} + b\,e^{3.5t},
\end{equation}
where $a$ and $b$ are coefficients to be determined empirically. 
Accordingly, $L_{\rm bol}^{\rm clump}$ is expected to evolve as a relatively flat function of $t$ during the early stages (small $t$), followed by a sharp increase at later times (large $t$) as the intrinsic stellar luminosity becomes dominant.

This predicted trend is indeed observed in the $L_{\rm bol}$–$t$ diagram for the ATLASGAL clumps (lower panel of Figure~\ref{fig_maxcore_L}), supporting the scenario of exponentially accelerating star formation. A fit to the observed data yields
\begin{equation}
    a = 40\,L_{\odot}, \qquad b = 10\,L_{\odot}.
\end{equation}
The intrinsic stellar luminosity dominates over the accretion luminosity from an early stage ($t \gtrsim 0.5\,t_0$), consistent with the rapid emergence of massive protostars in dense environments.

\begin{figure}
    \centering
    \includegraphics[width=0.99\linewidth]{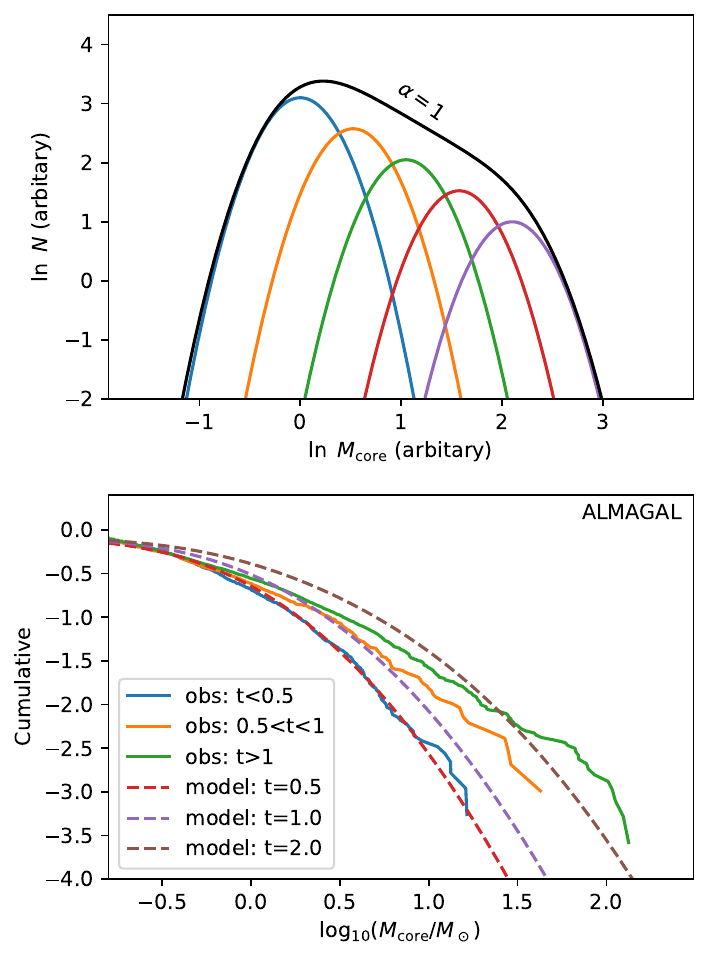}
    \caption{Upper: A schematic illustration showing how shifted ICMFs (colored curves) combine to produce a top-heavy CMF (black curve).  
The upper shifted ICMFs represent newly formed cores, while the accretion process shifts the mass function of older cores toward higher masses, forming right-shifted ICMFs.  
Lower: Cumulative functions of the observed CMFs for three time bins, together with the corresponding modeled CMFs (see upper panel), where the $x$-axis values are linearly scaled to match the observations.
    }
    \label{fig:cmf}
\end{figure}

\section{Discussion}
\subsection{CMF under accelerating scenario}\label{sec_cmf}
Assuming a constant ICMF with slope $\alpha_{\rm ICMF}>1$, Eqs.~\ref{eq_coregrow} and \ref{eq_Ngrow} yield
\begin{equation}
\begin{aligned}
    {\rm CMF}_{\rm ln}(\ln M_{\rm core},t) 
    &\propto \int_0^t e^\tau\, {\rm ICMF}_{\rm ln}\big(\ln M_{\rm core}-(t-\tau)\big)\, d\tau \\
    &\propto \Bigl(f(y;t) \,\mathop{\ast}\, {\rm ICMF}_{\rm ln}[y]\Bigr)(\ln M_{\rm core}),
\end{aligned}
 \label{eq_CMFln}
\end{equation}
where $f$ is a truncated exponential function,
\begin{equation}
    f(y;t) = 
    \begin{cases}
        e^{-y}, & 0<y<t, \\
        0, & \text{otherwise},
    \end{cases}
\end{equation}
and $\ast$ denotes the convolution operator.
On the linear scale, the CMF is related to its logarithmic counterpart via
\begin{equation}
    {\rm CMF}(M_{\rm core};t) = \frac{1}{M_{\rm core}}\,{\rm CMF}_{\rm ln}(\ln M_{\rm core};t),
\end{equation}
and similarly for the ICMF.  

Since the shallowest exponential tail is preserved under convolution \citep{2025arXiv250220458L}, the high-mass end of the CMF retains an approximately exponential form with truncation at the largest masses, independent of the detailed shape of the ICMF.  
That is, at the high-mass end, 
\begin{equation}
    {\rm CMF}_{\rm ln}(\ln M_{\rm core};t) \propto \exp(-\ln M_{\rm core}) \propto M_{\rm core}^{-1},
\end{equation}
yielding an effective slope $\alpha = 1$ regardless of the specific form of the ICMF.
The upper panel of Figure~\ref{fig:cmf} illustrates how the convolution process produces a progressively top-heavy CMF.
Note that the inferred value of the exponential index of the CMF still requires further observational testing \citep{2022A&A...664A..26P,2024A&A...690A..33L}, as it remains an open question.

In the lower panel of Figure~\ref{fig:cmf}, the cumulative function of ${\rm CMF}_{\rm ln}$ for the ALMAGAL samples is shown for three time bins.  
The CMF builds up and becomes flatter with evolution.  
This behavior is consistent with the trend reported by \citet{2025A&A...696A.151C}, who binned the samples by $L/M$, and is also qualitatively consistent with the model prediction given by Eq.~\ref{eq_CMFln} (see lower panel of Figure~\ref{fig:cmf}).  
The slightly enhanced number of high-mass cores, or the deficit of low-mass ones, may result from the limited sensitivity and field of view of the observations, which tend to filter out some of the low-mass cores, as well as from the specific form of the ICMF at the low-mass end, which remains poorly constrained at present.

\subsection{Star formation law}\label{sec_sf_law}

Star formation in dense molecular gas exhibits a remarkably tight correlation with the mass of dense gas, often referred to as the dense gas star formation law \citep{2004ApJ...606..271G}. Observations of individual molecular clouds \citep[e.g.,][]{2017A&A...604A..74S} as well as galaxy-scale statistics \citep[e.g.,][]{2004ApJ...606..271G} show that the star formation rate (SFR) scales approximately linearly with the mass of dense gas (traced by HCN or HCO$^+$),
\begin{equation}
    \mathrm{SFR} \propto M_{\rm dense},
\end{equation}
where $M_{\rm dense}$ denotes the mass of gas above a given density threshold. In individual molecular clouds, this dense gas is typically close to virial equilibrium, with $M_{\rm vir} \gtrsim M_{\rm gas}$.  
The cloud-scale SF law highlights a set of nearly universal properties of dense clumps within individual clouds. In this work, the exponentially accelerating mass assembly of dense clumps allows them to rapidly reach a saturated state of star formation, naturally leading to the observed correlation between dense gas mass and SFR.  

Importantly, the SF law can also be extended to the scale of an individual dense clump. In this case, the SFR scales with the mass of star-forming gas within the clump, $M_{\rm SF}$, which we identify with the virial mass of the clump,  
\begin{equation}
    \mathrm{SFR} \propto \dot{M}_{\rm SF} \propto M_{\rm SF} \propto M_{\rm vir} \propto e^t,
\end{equation}
reflecting the exponential growth of the star-forming mass over time in our model.  
This formulation suggests that the observed linear correlation between dense gas mass and SFR, whether measured for individual clouds or on galactic scales, can be understood as a natural consequence of the underlying clump-scale star formation process.

In extreme environments such as the Central Molecular Zone (CMZ), the SFR is observed to be significantly lower than the value predicted from the dense gas mass \citep[e.g.,][]{10.1093/mnras/sts376}, implying either a higher effective density threshold for star formation or a lower star formation efficiency per free-fall time ($\eta_{\rm ff}$), likely due to strong shear and tidal effects.  
We suggest that the virial mass of dense gas, rather than the total dense gas mass, may be a more appropriate tracer of the SFR. This may help reconcile the apparently low SFR in the CMZ, where the observed SFR is roughly an order of magnitude below the expected value.  
If the CMZ gas structures are treated as giant clumps with a typical virial parameter of $\alpha_{\rm vir} = M_{\rm vir}/M_{\rm gas} \sim 0.1$, then adopting $M_{\rm vir}$ instead of $M_{\rm gas}$ yields a more consistent estimate of the expected SFR in these regions.

\subsection{Physical plausibility}
The exponential growth of protocluster mass revealed in our analysis can be naturally understood in terms of simultaneous core formation and mass assembly within dense clumps, collectively producing an exponential increase of the protocluster mass. This picture corresponds to a clump-fed scenario, in which the pre-existing dense clump acts as a mass reservoir while the protocluster grows dynamically through internal core formation and accretion. Although the host clump may continue to accrete material from its surroundings, its overall mass growth rate is typically slower than the exponential growth of the embedded protocluster once star formation begins. This indicates that the accelerating star formation is primarily regulated by the internal dynamics of core formation and accretion, rather than by the global growth of the clump.

One may question whether such exponential accretion is physically realistic. This can be understood by considering the hierarchical structure of forming protoclusters: subclusters form and merge within the clump, rapidly increasing the number of dense cores. Globally, the protocluster-scale accretion remains below the free-fall rate of the clump, but anisotropic, filamentary, or streamer-like flows can channel material onto the protocluster at near free-fall velocities for localized portions of the mass. Such streams may be tidally regulated by the central protocluster, with their widths scaling approximately as \citep{10.1093/mnrasl/slt077,2025arXiv250918495L}
\begin{equation}
    w_{\rm stream} \propto M_{\rm SF}^{0.5},
\end{equation}
and the mass flux along each stream proportional to \(M_{\rm SF}\),
\begin{equation}
    \dot{M}_{\rm SF} \propto w_{\rm stream}^2 \propto M_{\rm SF}.
\end{equation}
In this way, the combination of hierarchical core formation and filamentary accretion naturally constructs an effectively exponential growth of the protocluster mass. While this may not capture all the detailed physics of clump-scale accretion, it demonstrates that exponentially accelerating accretion is physically plausible within a self-consistent framework.

\subsection{Multi-scale accelerating star formation}
This work suggests that star formation, especially in the massive regime, may exhibit accelerating behavior on both core and clump scales (from 0.01 to 1 pc). At the core scale, this corresponds to the growth of core mass during the accretion phase. At the clump scale, it reflects the increase of protocluster mass accompanied  by the consumption of clump mass. In this picture, stellar accretion and clump mass consumption proceed in a closely linked manner on a timescale of $\sim10^{5}$ yr, typical for massive star formation. Massive clumps therefore act as the fundamental units for the formation of stellar clusters.

On cloud scales, \cite{2000ApJ...540..255P,2018MNRAS.474.4818C} inferred a similar accelerating star formation pattern by accounting for pre-main-sequence stars in young clusters in nearby low-mass star-forming clouds (with typical sizes of several pc). These studies are based on stellar age distributions integrated over entire molecular clouds, probing timescales of $\sim10^{6}$ yr. In these cases, the pre-main-sequence stars have already dispersed from their natal clumps and therefore no longer trace the instantaneous structure of the parental dense gas. The inferred trends may therefore describe cloud-integrated star formation across multiple regions and evolutionary stages, reflecting different physical scales, timescales, and stellar mass regimes.

Despite these differences, it is interesting to ask whether the two results are related within a hierarchical view of star formation. In this case, the accelerating behavior seen on cloud scales may partly arise from the formation, growth, and temporal sequence of the basic building blocks, the star-forming clumps, within the parent cloud. Alternatively, the formation patterns in high-mass and low-mass star-forming regions may be different.
The expansion and merging of protoclusters beyond the clump scale may play an important role in regulating star formation in a cloud.
From this point of view, combining millimeter observations of dense gas and cores with optical and infrared observations of more evolved pre-main-sequence stars may help connect star formation across scales, from core ($\sim 0.01$ pc) through clump ($\sim 1$ pc) to cloud ($\sim 10$ pc). 
Whether this connection can be explained as a simple sum of clump evolution, or whether additional coupling between scales is required, remains an open question.

\section{Summary}
We investigate the time-dependent growth of dense clumps and their embedded protoclusters using ATLASGAL and ALMAGAL data. By linearizing clump evolution with dust temperature, we find that both protocluster mass and the most massive cores grow exponentially over comparable timescales. 
This accelerating growth naturally reproduces the observed trends in luminosity, the core mass function, and the mass of the most massive cores, as well as the dense gas star formation law. Physically, the exponential increase arises from simultaneous core formation and accretion, with hierarchical subcluster assembly and anisotropic inflows providing a plausible pathway for rapid protocluster growth, while the host clump itself evolves more slowly.

\begin{acknowledgement}
X.L. acknowledges the support of the Strategic Priority Research Program of the Chinese Academy of Sciences under Grant No. XDB0800303. We thank Dr. Jinhua He of Yunnan Observatory and Dr. Guangxing Li of the Southwest Institute for Astronomy Research (SWIFAR) at Yunnan University (YNU) for helpful discussions.
\end{acknowledgement}

\bibliography{AcceleratingSF}

@ARTICLE{1998A&A...336..150M,
       author = {{Motte}, F. and {Andre}, P. and {Neri}, R.},
        title = "{The initial conditions of star formation in the rho Ophiuchi main cloud: wide-field millimeter continuum mapping}",
      journal = {\aap},
         year = 1998,
        month = aug,
       volume = {336},
        pages = {150-172},
       adsurl = {https://ui.adsabs.harvard.edu/abs/1998A&A...336..150M}
}

@ARTICLE{1955ApJ...121..161S,
       author = {{Salpeter}, Edwin E.},
        title = "{The Luminosity Function and Stellar Evolution.}",
      journal = {\apj},
         year = 1955,
        month = jan,
       volume = {121},
        pages = {161},
          doi = {10.1086/145971},
       adsurl = {https://ui.adsabs.harvard.edu/abs/1955ApJ...121..161S}
}

@ARTICLE{1998ApJ...498..541K,
       author = {{Kennicutt}, Jr., Robert C.},
        title = "{The Global Schmidt Law in Star-forming Galaxies}",
      journal = {\apj},
         year = 1998,
        month = may,
       volume = {498},
       number = {2},
        pages = {541-552},
          doi = {10.1086/305588},
archivePrefix = {arXiv},
       eprint = {astro-ph/9712213},
 primaryClass = {astro-ph},
       adsurl = {https://ui.adsabs.harvard.edu/abs/1998ApJ...498..541K}
}

@ARTICLE{2004ApJ...606..271G,
       author = {{Gao}, Yu and {Solomon}, Philip M.},
        title = "{The Star Formation Rate and Dense Molecular Gas in Galaxies}",
      journal = {\apj},
         year = 2004,
        month = may,
       volume = {606},
       number = {1},
        pages = {271-290},
          doi = {10.1086/382999},
archivePrefix = {arXiv},
       eprint = {astro-ph/0310339},
 primaryClass = {astro-ph},
       adsurl = {https://ui.adsabs.harvard.edu/abs/2004ApJ...606..271G}
}

@ARTICLE{
2003ApJ...585..850M,
       author = {{McKee}, Christopher F. and {Tan}, Jonathan C.},
        title = "{The Formation of Massive Stars from Turbulent Cores}",
      journal = {\apj},
         year = 2003,
        month = mar,
       volume = {585},
       number = {2},
        pages = {850-871},
          doi = {10.1086/346149},
archivePrefix = {arXiv},
       eprint = {astro-ph/0206037},
 primaryClass = {astro-ph},
       adsurl = {https://ui.adsabs.harvard.edu/abs/2003ApJ...585..850M}
}

@ARTICLE{2004MNRAS.349..735B,
       author = {{Bonnell}, Ian A. and {Vine}, Stephen G. and {Bate}, Matthew R.},
        title = "{Massive star formation: nurture, not nature}",
      journal = {\mnras},
         year = 2004,
        month = apr,
       volume = {349},
       number = {2},
        pages = {735-741},
          doi = {10.1111/j.1365-2966.2004.07543.x},
archivePrefix = {arXiv},
       eprint = {astro-ph/0401059},
 primaryClass = {astro-ph},
       adsurl = {https://ui.adsabs.harvard.edu/abs/2004MNRAS.349..735B}
}

@ARTICLE{2025RAA....25b5020L,
       author = {{Liu}, Xunchuan and {Liu}, Tie and {Mai}, Xiaofeng and {Cheng}, Yu and {Jiao}, Sihan and {Jiao}, Wenyu and {Liu}, Hongli and {Zhang}, Siju},
        title = "{Core Mass Function in View of Fractal and Turbulent Filaments and Fibers}",
      journal = {Research in Astronomy and Astrophysics},
         year = 2025,
        month = feb,
       volume = {25},
       number = {2},
          eid = {025020},
        pages = {025020},
          doi = {10.1088/1674-4527/adb15a},
archivePrefix = {arXiv},
       eprint = {2501.17502},
 primaryClass = {astro-ph.GA},
       adsurl = {https://ui.adsabs.harvard.edu/abs/2025RAA....25b5020L}
}

@ARTICLE{2016A&A...588A..29H,
       author = {{Heyer}, M. and {Gutermuth}, R. and {Urquhart}, J.~S. and {Csengeri}, T. and {Wienen}, M. and {Leurini}, S. and {Menten}, K. and {Wyrowski}, F.},
        title = "{The rate and latency of star formation in dense, massive clumps in the Milky Way}",
      journal = {\aap},
         year = 2016,
        month = apr,
       volume = {588},
          eid = {A29},
        pages = {A29},
          doi = {10.1051/0004-6361/201527681},
archivePrefix = {arXiv},
       eprint = {1601.06875},
 primaryClass = {astro-ph.GA},
       adsurl = {https://ui.adsabs.harvard.edu/abs/2016A&A...588A..29H}
}

@ARTICLE{2025A&A...696A.149M,
       author = {{Molinari}, S. and {Schilke}, P. and {Battersby}, C. and {Ho}, P.~T.~P. and {S{\'a}nchez-Monge}, {\'A}. and {Traficante}, A. and {Jones}, B. and {Beltr{\'a}n}, M.~T. and {Beuther}, H. and {Fuller}, G.~A. and {Zhang}, Q. and {Klessen}, R.~S. and {Walch}, S. and {Tang}, Y.-W. and {Benedettini}, M. and {Elia}, D. and {Coletta}, A. and {Mininni}, C. and {Schisano}, E. and {Avison}, A. and {Law}, C.~Y. and {Nucara}, A. and {Soler}, J.~D. and {Stroud}, G. and {Wallace}, J. and {Wells}, M.~R.~A. and {Ahmadi}, A. and {Brogan}, C.~L. and {Hunter}, T.~R. and {Liu}, S.-Y. and {Pezzuto}, S. and {Su}, Y.-N. and {Zimmermann}, B. and {Zhang}, T. and {Wyrowski}, F. and {De Angelis}, F. and {Liu}, S. and {Clarke}, S.~D. and {Fontani}, F. and {Klaassen}, P.~D. and {Koch}, P. and {Johnston}, K.~G. and {Lebreuilly}, U. and {Liu}, T. and {Lumsden}, S.~L. and {Moeller}, T. and {Moscadelli}, L. and {Kuiper}, R. and {Lis}, D. and {Peretto}, N. and {Pfalzner}, S. and {Rigby}, A.~J. and {Sanhueza}, P. and {Rygl}, K.~L.~J. and {van der Tak}, F. and {Zinnecker}, H. and {Amaral}, F. and {Bally}, J. and {Bronfman}, L. and {Cesaroni}, R. and {Goh}, K. and {Hoare}, M.~G. and {Hatchfield}, P. and {Hennebelle}, P. and {Henning}, T. and {Kim}, K.-T. and {Kim}, W.-J. and {Maud}, L. and {Merello}, M. and {Nakamura}, F. and {Plume}, R. and {Qin}, S.-L. and {Svoboda}, B. and {Testi}, L. and {Veena}, V.~S. and {Walker}, D.},
        title = "{ALMAGAL: I. The ALMA evolutionary study of high-mass protocluster formation in the Galaxy: Presentation of the survey and early results}",
      journal = {\aap},
         year = 2025,
        month = apr,
       volume = {696},
          eid = {A149},
        pages = {A149},
          doi = {10.1051/0004-6361/202452702},
archivePrefix = {arXiv},
       eprint = {2503.05555},
 primaryClass = {astro-ph.GA},
       adsurl = {https://ui.adsabs.harvard.edu/abs/2025A&A...696A.149M}
}

@ARTICLE{2009A&A...504..415S,
       author = {{Schuller}, F. and {Menten}, K.~M. and {Contreras}, Y. and {Wyrowski}, F. and {Schilke}, P. and {Bronfman}, L. and {Henning}, T. and {Walmsley}, C.~M. and {Beuther}, H. and {Bontemps}, S. and {Cesaroni}, R. and {Deharveng}, L. and {Garay}, G. and {Herpin}, F. and {Lefloch}, B. and {Linz}, H. and {Mardones}, D. and {Minier}, V. and {Molinari}, S. and {Motte}, F. and {Nyman}, L.-{\r{A}}. and {Reveret}, V. and {Risacher}, C. and {Russeil}, D. and {Schneider}, N. and {Testi}, L. and {Troost}, T. and {Vasyunina}, T. and {Wienen}, M. and {Zavagno}, A. and {Kovacs}, A. and {Kreysa}, E. and {Siringo}, G. and {Wei{\ss}}, A.},
        title = "{ATLASGAL - The APEX telescope large area survey of the galaxy at 870 {\ensuremath{\mu}}m}",
      journal = {\aap},
         year = 2009,
        month = sep,
       volume = {504},
       number = {2},
        pages = {415-427},
          doi = {10.1051/0004-6361/200811568},
archivePrefix = {arXiv},
       eprint = {0903.1369},
 primaryClass = {astro-ph.GA},
       adsurl = {https://ui.adsabs.harvard.edu/abs/2009A&A...504..415S}
}

@ARTICLE{2024RAA....24g5001L,
       author = {{Liu}, Mingchao and {He}, Jinhua and {Guo}, Zhen and {Ge}, Jixing and {Tang}, Yuping},
        title = "{Can near-to-mid Infrared Spectral Energy Distribution Quantitatively Trace Protoplanetary Disk Evolution?}",
      journal = {Research in Astronomy and Astrophysics},
         year = 2024,
        month = jul,
       volume = {24},
       number = {7},
          eid = {075001},
        pages = {075001},
          doi = {10.1088/1674-4527/ad4b5c},
archivePrefix = {arXiv},
       eprint = {2404.11048},
 primaryClass = {astro-ph.SR},
       adsurl = {https://ui.adsabs.harvard.edu/abs/2024RAA....24g5001L}
}

@ARTICLE{2024RAA....24b5009L,
       author = {{Liu}, Xunchuan and {Liu}, Tie and {Zhu}, Lei and {Garay}, Guido and {Liu}, Hong-Li and {Goldsmith}, Paul and {Evans}, Neal and {Kim}, Kee-Tae and {Liu}, Sheng-Yuan and {Xu}, Fengwei and {Lu}, Xing and {Tej}, Anandmayee and {Mai}, Xiaofeng and {Bronfman}, Leonardo and {Li}, Shanghuo and {Mardones}, Diego and {Stutz}, Amelia and {Tatematsu}, Ken'ichi and {Wang}, Ke and {Zhang}, Qizhou and {Qin}, Sheng-Li and {Zhou}, Jianwen and {Luo}, Qiuyi and {Zhang}, Siju and {Cheng}, Yu and {He}, Jinhua and {Gu}, Qilao and {Li}, Ziyang and {Zhang}, Zhenying and {Zhang}, Suinan and {Saha}, Anindya and {Dewangan}, Lokesh and {Sanhueza}, Patricio and {Shen}, Zhiqiang},
        title = "{The ALMA-QUARKS Survey. I. Survey Description and Data Reduction}",
      journal = {Research in Astronomy and Astrophysics},
         year = 2024,
        month = feb,
       volume = {24},
       number = {2},
          eid = {025009},
        pages = {025009},
          doi = {10.1088/1674-4527/ad0d5c},
archivePrefix = {arXiv},
       eprint = {2311.08651},
 primaryClass = {astro-ph.GA},
       adsurl = {https://ui.adsabs.harvard.edu/abs/2024RAA....24b5009L}
}

@ARTICLE{2018MNRAS.473.1059U,
       author = {{Urquhart}, J.~S. and {K{\"o}nig}, C. and {Giannetti}, A. and {Leurini}, S. and {Moore}, T.~J.~T. and {Eden}, D.~J. and {Pillai}, T. and {Thompson}, M.~A. and {Braiding}, C. and {Burton}, M.~G. and {Csengeri}, T. and {Dempsey}, J.~T. and {Figura}, C. and {Froebrich}, D. and {Menten}, K.~M. and {Schuller}, F. and {Smith}, M.~D. and {Wyrowski}, F.},
        title = "{ATLASGAL - properties of a complete sample of Galactic clumps}",
      journal = {\mnras},
         year = 2018,
        month = jan,
       volume = {473},
       number = {1},
        pages = {1059-1102},
          doi = {10.1093/mnras/stx2258},
archivePrefix = {arXiv},
       eprint = {1709.00392},
 primaryClass = {astro-ph.GA},
       adsurl = {https://ui.adsabs.harvard.edu/abs/2018MNRAS.473.1059U}
}

@ARTICLE{2022MNRAS.510.3389U,
       author = {{Urquhart}, J.~S. and {Wells}, M.~R.~A. and {Pillai}, T. and {Leurini}, S. and {Giannetti}, A. and {Moore}, T.~J.~T. and {Thompson}, M.~A. and {Figura}, C. and {Colombo}, D. and {Yang}, A.~Y. and {K{\"o}nig}, C. and {Wyrowski}, F. and {Menten}, K.~M. and {Rigby}, A.~J. and {Eden}, D.~J. and {Ragan}, S.~E.},
        title = "{ATLASGAL - evolutionary trends in high-mass star formation}",
      journal = {\mnras},
         year = 2022,
        month = mar,
       volume = {510},
       number = {3},
        pages = {3389-3407},
          doi = {10.1093/mnras/stab3511},
archivePrefix = {arXiv},
       eprint = {2111.12816},
 primaryClass = {astro-ph.GA},
       adsurl = {https://ui.adsabs.harvard.edu/abs/2022MNRAS.510.3389U}
}

@ARTICLE{2025A&A...696A.151C,
       author = {{Coletta}, A. and {Molinari}, S. and {Schisano}, E. and {Traficante}, A. and {Elia}, D. and {Benedettini}, M. and {Mininni}, C. and {Soler}, J.~D. and {S{\'a}nchez-Monge}, {\'A}. and {Schilke}, P. and {Battersby}, C. and {Fuller}, G.~A. and {Beuther}, H. and {Zhang}, Q. and {Beltr{\'a}n}, M.~T. and {Jones}, B. and {Klessen}, R.~S. and {Walch}, S. and {Fontani}, F. and {Avison}, A. and {Brogan}, C.~L. and {Clarke}, S.~D. and {Hatchfield}, P. and {Hennebelle}, P. and {Ho}, P.~T.~P. and {Hunter}, T.~R. and {Johnston}, K.~G. and {Klaassen}, P.~D. and {Koch}, P.~M. and {Kuiper}, R. and {Lis}, D.~C. and {Liu}, T. and {Lumsden}, S.~L. and {Maruccia}, Y. and {M{\"o}ller}, T. and {Moscadelli}, L. and {Nucara}, A. and {Rigby}, A.~J. and {Rygl}, K.~L.~J. and {Sanhueza}, P. and {van der Tak}, F. and {Wells}, M.~R.~A. and {Wyrowski}, F. and {De Angelis}, F. and {Liu}, S. and {Ahmadi}, A. and {Bronfman}, L. and {Liu}, S.-Y. and {Su}, Y.-N. and {Tang}, Y. and {Testi}, L. and {Zinnecker}, H.},
        title = "{ALMAGAL: III. Compact source catalog: Fragmentation statistics and physical evolution of the core population}",
      journal = {\aap},
         year = 2025,
        month = apr,
       volume = {696},
          eid = {A151},
        pages = {A151},
          doi = {10.1051/0004-6361/202452706},
archivePrefix = {arXiv},
       eprint = {2503.05663},
 primaryClass = {astro-ph.GA},
       adsurl = {https://ui.adsabs.harvard.edu/abs/2025A&A...696A.151C}
}

@ARTICLE{2022A&A...664A..26P,
       author = {{Pouteau}, Y. and {Motte}, F. and {Nony}, T. and {Galv{\'a}n-Madrid}, R. and {Men'shchikov}, A. and {Bontemps}, S. and {Robitaille}, J.-F. and {Louvet}, F. and {Ginsburg}, A. and {Herpin}, F. and {L{\'o}pez-Sepulcre}, A. and {Dell'Ova}, P. and {Gusdorf}, A. and {Sanhueza}, P. and {Stutz}, A.~M. and {Brouillet}, N. and {Thomasson}, B. and {Armante}, M. and {Baug}, T. and {Bonfand}, M. and {Busquet}, G. and {Csengeri}, T. and {Cunningham}, N. and {Fern{\'a}ndez-L{\'o}pez}, M. and {Liu}, H.-L. and {Olguin}, F. and {Towner}, A.~P.~M. and {Bally}, J. and {Braine}, J. and {Bronfman}, L. and {Joncour}, I. and {Gonz{\'a}lez}, M. and {Hennebelle}, P. and {Lu}, X. and {Menten}, K.~M. and {Moraux}, E. and {Tatematsu}, K. and {Walker}, D. and {Whitworth}, A.~P.},
        title = "{ALMA-IMF. III. Investigating the origin of stellar masses: top-heavy core mass function in the W43-MM2\&MM3 mini-starburst}",
      journal = {\aap},
         year = 2022,
        month = aug,
       volume = {664},
          eid = {A26},
        pages = {A26},
          doi = {10.1051/0004-6361/202142951},
archivePrefix = {arXiv},
       eprint = {2203.03276},
 primaryClass = {astro-ph.GA},
       adsurl = {https://ui.adsabs.harvard.edu/abs/2022A&A...664A..26P}
}

@ARTICLE{2018ApJ...853..160C,
       author = {{Cheng}, Yu and {Tan}, Jonathan C. and {Liu}, Mengyao and {Kong}, Shuo and {Lim}, Wanggi and {Andersen}, Morten and {Da Rio}, Nicola},
        title = "{The Core Mass Function in the Massive Protocluster G286.21+0.17 Revealed by ALMA}",
      journal = {\apj},
         year = 2018,
        month = feb,
       volume = {853},
       number = {2},
          eid = {160},
        pages = {160},
          doi = {10.3847/1538-4357/aaa3f1},
archivePrefix = {arXiv},
       eprint = {1706.06584},
 primaryClass = {astro-ph.GA},
       adsurl = {https://ui.adsabs.harvard.edu/abs/2018ApJ...853..160C}
}

@ARTICLE{2021ApJ...921...48S,
       author = {{Su{\'a}rez}, Genaro and {Galv{\'a}n-Madrid}, Roberto and {Aguilar}, Luis and {Ginsburg}, Adam and {Srinivasan}, Sundar and {Liu}, Hauyu Baobab and {Rom{\'a}n-Z{\'u}{\~n}iga}, Carlos G.},
        title = "{A Core Mass Function Indistinguishable from the Salpeter Stellar Initial Mass Function Using 1000 au Resolution ALMA Observations}",
      journal = {\apj},
         year = 2021,
        month = nov,
       volume = {921},
       number = {1},
          eid = {48},
        pages = {48},
          doi = {10.3847/1538-4357/ac1bb9},
archivePrefix = {arXiv},
       eprint = {2107.14288},
 primaryClass = {astro-ph.GA},
       adsurl = {https://ui.adsabs.harvard.edu/abs/2021ApJ...921...48S}
}

@ARTICLE{2024A&A...690A..33L,
       author = {{Louvet}, F. and {Sanhueza}, P. and {Stutz}, A. and {Men'shchikov}, A. and {Motte}, F. and {Galv{\'a}n-Madrid}, R. and {Bontemps}, S. and {Pouteau}, Y. and {Ginsburg}, A. and {Csengeri}, T. and {Di Francesco}, J. and {Dell'Ova}, P. and {Gonz{\'a}lez}, M. and {Didelon}, P. and {Braine}, J. and {Cunningham}, N. and {Thomasson}, B. and {Lesaffre}, P. and {Hennebelle}, P. and {Bonfand}, M. and {Gusdorf}, A. and {{\'A}lvarez-Guti{\'e}rrez}, R.~H. and {Nony}, T. and {Busquet}, G. and {Olguin}, F. and {Bronfman}, L. and {Salinas}, J. and {Fernandez-Lopez}, M. and {Moraux}, E. and {Liu}, H.~L. and {Lu}, X. and {Huei-Ru}, V. and {Towner}, A. and {Valeille-Manet}, M. and {Brouillet}, N. and {Herpin}, F. and {Lefloch}, B. and {Baug}, T. and {Maud}, L. and {L{\'o}pez-Sepulcre}, A. and {Svoboda}, B.},
        title = "{ALMA-IMF: XV. Core mass function in the high-mass star formation regime}",
      journal = {\aap},
         year = 2024,
        month = oct,
       volume = {690},
          eid = {A33},
        pages = {A33},
          doi = {10.1051/0004-6361/202345986},
archivePrefix = {arXiv},
       eprint = {2407.18719},
 primaryClass = {astro-ph.GA},
       adsurl = {https://ui.adsabs.harvard.edu/abs/2024A&A...690A..33L}
}

@ARTICLE{2001MNRAS.323..785B,
       author = {{Bonnell}, I.~A. and {Bate}, M.~R. and {Clarke}, C.~J. and {Pringle}, J.~E.},
        title = "{Competitive accretion in embedded stellar clusters}",
      journal = {\mnras},
         year = 2001,
        month = may,
       volume = {323},
       number = {4},
        pages = {785-794},
          doi = {10.1046/j.1365-8711.2001.04270.x},
archivePrefix = {arXiv},
       eprint = {astro-ph/0102074},
 primaryClass = {astro-ph},
       adsurl = {https://ui.adsabs.harvard.edu/abs/2001MNRAS.323..785B}
}

@ARTICLE{1991ApJ...375..288P,
       author = {{Palla}, Francesco and {Stahler}, Steven W.},
        title = "{The Evolution of Intermediate-Mass Protostars. I. Basic Results}",
      journal = {\apj},
         year = 1991,
        month = jul,
       volume = {375},
        pages = {288},
          doi = {10.1086/170188},
       adsurl = {https://ui.adsabs.harvard.edu/abs/1991ApJ...375..288P}
}

@ARTICLE{2010ApJ...721..478H,
       author = {{Hosokawa}, Takashi and {Yorke}, Harold W. and {Omukai}, Kazuyuki},
        title = "{Evolution of Massive Protostars Via Disk Accretion}",
      journal = {\apj},
         year = 2010,
        month = sep,
       volume = {721},
       number = {1},
        pages = {478-492},
          doi = {10.1088/0004-637X/721/1/478},
archivePrefix = {arXiv},
       eprint = {1005.2827},
 primaryClass = {astro-ph.SR},
       adsurl = {https://ui.adsabs.harvard.edu/abs/2010ApJ...721..478H}
}

@ARTICLE{2025arXiv250220458L,
       author = {{Liu}, Xunchuan},
        title = "{Turbulence in virtual: Origin of the variance and skewness of density function}",
      journal = {arXiv e-prints},
         year = 2025,
        month = feb,
          eid = {arXiv:2502.20458},
        pages = {arXiv:2502.20458},
          doi = {10.48550/arXiv.2502.20458},
archivePrefix = {arXiv},
       eprint = {2502.20458},
 primaryClass = {astro-ph.GA},
       adsurl = {https://ui.adsabs.harvard.edu/abs/2025arXiv250220458L}
}

@article{10.1093/mnras/sts376,
    author = {Longmore, S. N. and Bally, J. and Testi, L. and Purcell, C. R. and Walsh, A. J. and Bressert, E. and Pestalozzi, M. and Molinari, S. and Ott, J. and Cortese, L. and Battersby, C. and Murray, N. and Lee, E. and Kruijssen, J. M. D. and Schisano, E. and Elia, D.},
    title = {Variations in the Galactic star formation rate and density thresholds for star formation},
    journal = {Monthly Notices of the Royal Astronomical Society},
    volume = {429},
    number = {2},
    pages = {987-1000},
    year = {2012},
    month = {12},
    issn = {0035-8711},
    doi = {10.1093/mnras/sts376},
    url = {https://doi.org/10.1093/mnras/sts376},
    eprint = {https://academic.oup.com/mnras/article-pdf/429/2/987/18448189/sts376.pdf},
}

@ARTICLE{2017A&A...604A..74S,
       author = {{Shimajiri}, Y. and {Andr{\'e}}, Ph. and {Braine}, J. and {K{\"o}nyves}, V. and {Schneider}, N. and {Bontemps}, S. and {Ladjelate}, B. and {Roy}, A. and {Gao}, Y. and {Chen}, H.},
        title = "{Testing the universality of the star-formation efficiency in dense molecular gas}",
      journal = {\aap},
         year = 2017,
        month = aug,
       volume = {604},
          eid = {A74},
        pages = {A74},
          doi = {10.1051/0004-6361/201730633},
archivePrefix = {arXiv},
       eprint = {1705.00213},
 primaryClass = {astro-ph.GA},
       adsurl = {https://ui.adsabs.harvard.edu/abs/2017A&A...604A..74S}
}

@ARTICLE{2025arXiv250918495L,
       author = {{Li}, Guang-Xing},
        title = "{Tidally-Controlled Fragmentation around Black Holes, Massive Clumps, Protostars, and the Galactic Center}",
      journal = {arXiv e-prints},
         year = 2025,
        month = sep,
          eid = {arXiv:2509.18495},
        pages = {arXiv:2509.18495},
          doi = {10.48550/arXiv.2509.18495},
archivePrefix = {arXiv},
       eprint = {2509.18495},
 primaryClass = {astro-ph.GA},
       adsurl = {https://ui.adsabs.harvard.edu/abs/2025arXiv250918495L}
}

@article{10.1093/mnrasl/slt077,
    author = {Jog, Chanda J.},
    title = {Jeans instability criterion modified by external tidal field},
    journal = {Monthly Notices of the Royal Astronomical Society: Letters},
    volume = {434},
    number = {1},
    pages = {L56-L60},
    year = {2013},
    month = {06},
    issn = {1745-3925},
    doi = {10.1093/mnrasl/slt077},
    url = {https://doi.org/10.1093/mnrasl/slt077},
    eprint = {https://academic.oup.com/mnrasl/article-pdf/434/1/L56/54658158/mnrasl_434_1_l56.pdf},
}

@ARTICLE{2000ApJ...540..255P,
       author = {{Palla}, Francesco and {Stahler}, Steven W.},
        title = "{Accelerating Star Formation in Clusters and Associations}",
      journal = {\apj},
         year = 2000,
        month = sep,
       volume = {540},
       number = {1},
        pages = {255-270},
          doi = {10.1086/309312},
       adsurl = {https://ui.adsabs.harvard.edu/abs/2000ApJ...540..255P}
}

@ARTICLE{2018MNRAS.474.4818C,
       author = {{Caldwell}, Spencer and {Chang}, Philip},
        title = "{The accelerating pace of star formation}",
      journal = {\mnras},
         year = 2018,
        month = mar,
       volume = {474},
       number = {4},
        pages = {4818-4823},
          doi = {10.1093/mnras/stx3037},
archivePrefix = {arXiv},
       eprint = {1711.07512},
 primaryClass = {astro-ph.GA},
       adsurl = {https://ui.adsabs.harvard.edu/abs/2018MNRAS.474.4818C}
}
\bibliographystyle{aa}

\end{document}